\documentclass[iop]{emulateapj}
\newcommand{\actaa}{Acta Astron.}

\shorttitle{Inferring mode inertias in evolved solar-like stars}
\shortauthors{Benomar et al.}

\begin{document}

%% LaTeX will automatically break titles if they run longer than
%% one line. However, you may use \\ to force a line break if
%% you desire.

\title{Asteroseismology of evolved stars with Kepler: a new way to constrain stellar interiors using mode inertias}

\author{O. Benomar\altaffilmark{1,6, 7}, K. Belkacem\altaffilmark{2}, T.R. Bedding\altaffilmark{1,6},  D. Stello\altaffilmark{1,6}, M.P. Di Mauro\altaffilmark{4}, R. Ventura\altaffilmark{5} , B. Mosser\altaffilmark{2}, M.J. Goupil\altaffilmark{2}, R. Samadi\altaffilmark{2}, R.A. Garcia\altaffilmark{3}} %, J. Ballot\altaffilmark{6}}

\altaffiltext{1}{Sydney Institute for Astronomy (SIfA), School of Physics,
  University of Sydney, NSW 2006, Australia} 
\altaffiltext{2}{LESIA, Observatoire de Paris, CNRS UMR 8109, Universit\'e Paris Diderot, 5 place J. Janssen, 92195 Meudon, France}
\altaffiltext{3}{{Laboratoire AIM, CEA/DSM-CNRS-Universit\'e Paris Diderot; IRFU/SAp, Centre de Saclay, 91191 Gif-sur-Yvette Cedex, France}}
\altaffiltext{4}{ INAF-IAPS Istituto di Astrofisica e Planetologia Spaziali\\ Via del Fosso del Cavaliere 100, 00133, Roma, Italy}
\altaffiltext{5}{INAF-Astrophyscial Observatory of Catania, Via S. Sofia 78, 95123, Catania, Italy}
\altaffiltext{6}{Stellar Astrophysics Centre, Department of Physics and Astronomy, Aarhus University, Ny Munkegade 120, DK-8000 Aarhus C, Denmark}
\altaffiltext{7}{Department of Astronomy, The University of Tokyo, 113-033, Japan}

\begin{abstract}
 Asteroseismology of evolved solar-like stars is experiencing a growing interest due to the wealth of observational data from space-borne instruments such as the \emph{CoRoT} and \emph{Kepler} spacecraft. In particular, the recent detection of mixed modes, which probe both the innermost and uppermost layers of stars, paves the way for inferring the internal structure of stars along their evolution through the subgiant and red giant phases. Mixed modes can also place stringent constraints on the physics of such stars and on their global properties (mass, age, etc...). Here, using two \emph{Kepler} stars (KIC 4351319 and KIC 6442183), we demonstrate that measurements of mixed mode characteristics allow us to estimate the mode inertias, providing a new and additional diagnostics on the mode trapping and subsequently on the internal structure of evolved stars. We however stress that the accuracy may be sensitive to non-adiabatic effects.
\end{abstract}

\keywords{stars: oscillations, stars: interiors, methods: data analysis}

\section{Introduction}\label{sec:1}

Asteroseismology, with the help of the space-borne missions such as CoRoT  \citep[e.g.][]{Baglin2006a,Baglin2006b,Michel2008} and \emph{Kepler}  \citep[e.g.][]{Borucki2010} gives us access to the internal structure of stars. This is made possible through the detection of oscillation modes propagating throughout the stars \citep{Appourchaux2008, Metcalfe2012, Gizon2013}.

While for main sequence stars we observe only accoustic modes, subgiants and red giants also show a rich spectrum of so-called \emph{mixed modes} \citep[e.g.][]{Beck2011Science,Bedding2011Nature,Mosser2011a, Benomar2013a}. 
They have a dual nature, behaving as acoustic modes in the stellar envelope and as gravity modes in the core. They are therefore detectable at the stellar surface while providing information on the innermost regions. This aspect of mixed modes has motivated extensive theoretical work \citep[e.g.][]{Dziembowski1971,Scuflaire1974,Osaki1975,Aizenman1977,Dziembowski2001,JCD2004,Dupret2009}.

Oscillations are very sensitive diagnostics that can be used to constrain the stellar structure \emph{and to} give an insight to internal properties of stars, such as the extent of the convective zone \citep{Roxburgh2009}.
In particular, mixed modes provide valuable information on the core structure which, for example, allows us to better constrain the evolutionary status of evolved low-mass stars  \citep{Bedding2011Nature,Mosser2011a}. The strength of mode mixing depends mainly on the trapping of the mode and therefore on the evanescent region between the inner and outer resonant cavities of stars.  
%\textbf{However, even if frequencies, or frequency ratios are traditionnally used to probe stellar interiors, other mode properties can bring additionnal stringent constraints. Here, we show that by adding an extra dimension, mode inertias, conjugated with mode frequencies, could help us to infer more precisely stellar interiors.}

To date, asteroseismology has generally used the mode {\it frequencies} to probe stellar interiors. In this \emph{Letter}, we show that we can also use mode {\it inertias}. Theoretical calculations \citep[e.g.][]{Dupret2009} show that inertias of mixed modes vary with frequency in a way that is related to the efficiency of the trapping, which changes significantly as the star evolves. Moreover, \cite{Samadi2008} showed that mode inertia depends strongly on turbulent pressure.
Hence, mode inertias can provide stringent constraints on mode trapping, reveal stellar interiors and potentially provide a new insight on the physics of stars. Furthermore, based on an asymptotic analysis, \cite{Goupil2013} proposed a way to observationally derive the ratio between the inertia in the $g$ cavity and the total inertia of a mixed mode.

Motivated by the latest \emph{Kepler} observations, and based on the mode fitting results from \cite{Benomar2013a}, we show that in conjunction with frequencies, mode inertias could help us to infer precisely the cavities of oscillations. %stellar interiors \textbf{, such as the depth of the convective envelope, the size of the core or the internal rotation profile}. 
We describe in Section \ref{sec:2} a method for inferring mode inertias from the measurements of mode amplitudes and linewidths. In Section \ref{sec:3} we apply the procedure to two stars observed by \emph{Kepler}, KIC 6442183 and KIC 4351319, and compare these results with models. We present diagnostics of the stellar structure of evolved stars, derived from the measurements of inertia in Section \ref{sec:4}. %Finally, Section \ref{sec:5} is dedicated to the discussion and conclusion.

\section{Inference of mode inertia ratios} \label{sec:2}

\subsection{Preliminary definitions}
    
In this section we show how to use observed mode amplitudes and linewidths to measure mode inertia for stars showing mixed modes. Let us consider a stochastically-driven mode. If this mode is resolved, \emph{i.e.} if the duration of the observations is significantly longer than the mode lifetime, it has a Lorentzian profile in the power spectrum \citep[e.g. ][]{Appourchaux2012}. In contrast, if the duration of the observations is shorter than the mode lifetime, the mode will have an unresolved sinc-squared profile \citep[such as expected for solar $g$ modes, see][]{Belkacem2009}. Thus, it is only possible to measure linewidths for resolved modes and we restrict our discussion to this case.

The height $H$ of a mode is given by \citep[e.g.][]{Libbrecht1988,Chaplin1998,Baudin2005,Chaplin2005,Belkacem2006b}
\begin{equation}
\label{def:hauteur}
	H = \frac{P}{2 \eta^2 \mathcal{M}} \, , 
\end{equation}
where  $P$ is the excitation rate, $\eta$ is the damping rate, and $\mathcal{M}$ is the mode mass, which is proportional to the mode inertia $I$ \citep[e.g. ][]{GMK94}:
\begin{equation} \label{def:Mode_mass_inertia}
	\mathcal{M} =  \frac{I}{\left| \vec \xi (R) \right|^2}\,, \quad {\rm with} \quad I= \int_0^M  \left| \vec{\xi}^2 \right| {\rm d}m \,.
\end{equation}
%where $\vert  \; \vert$ is the norm,  
%where $\vec \xi (R)$, $R$, ${\rm d} m$, $M$ are the displacement at the photosphere, the stellar radius, a mass element and the mass of the star, respectively. 
Here, $\vec \xi (R)$, $R$, ${\rm d} m$, $M$ are the displacement vector, the stellar radius, the mass element and the mass of the star, respectively. 

Power spectrum analysis mostly relies on fitting Lorentzian profiles to the observed modes. The usual form is
\begin{equation}
	L(\nu) = \frac{H}{1 + 4(\nu - \nu_c)^2/\Gamma^2},
\end{equation}
such that $\Gamma = \eta/\pi$ is the full-width-at-half-maximum (hereafter called linewidth) %(referred in the following as linewidth) 
and $\nu_c$ is the central frequency. The observables needed to estimate the mode inertia are the height $H$ and the linewidth $\Gamma$. Note that they are also related to the mode amplitude: 
\begin{equation}
\label{def:amplitude}
	A^2 = \pi H \, \Gamma \ /2. 
\end{equation}

%\begin{figure}[]
%	\includegraphics[angle=180,totalheight=6.cm]{Figures/plot_othman_HR.eps}
%	\caption{Position in the HR-Diagram of the two modeled stars, on a $1 M_\odot$ track. \textbf{Model A refers to the subgiant, Model B to the red giant}.}
%	\label{fig:HRdiagram}
%\end{figure}

\subsection{Relation between mode inertia, amplitude and linewidth}

%Most recent observations from CoRoT and \emph{Kepler} revealed resolved mixed modes in subgiants and red giants such that the aforementioned considerations are valid. Even more

We recently used \emph{Kepler} data to accurately measure the Lorentzian profiles of oscillation modes \citep{Benomar2013a}. 
%, paving the way to a seismic inference of the internal structure of evolved solar-like stars. 
To go a step further, and guided by those results, we now consider the comparison of two neighboring modes: 
\begin{enumerate}
\item a radial mode ($\ell=0$), denoted by the subscript $0$, and
\item a dipole mode ($\ell=1$) that exhibits a mixed character, i.e., pressure-like behavior in the envelope and gravity-like behavior in the innermost layers. The subscript $1$ will be used for this mode.
\end{enumerate}
We choose modes that are resolved, allowing us to apply Equation~\ref{def:hauteur} for computing their heights and assume modes of neighbor frequency ($\nu_0$ and $\nu_1$). %We also choose modes whose frequencies (namely $\nu_0$ and $\nu_1$) are  close to each other. %More precisely, we require that $\vert \nu_0-\nu_1 \vert < \Delta\nu/2$, where $\Delta \nu$ is the large separation.  This allows us to assume the driving of both modes to be the same. 
%\end{enumerate}

From  Equation~\ref{def:hauteur} and the fact that $\Gamma=\eta/\pi$, the ratio of mode heights can be written  
\begin{equation}
\label{rapport_hauteurs}
\frac{H_0}{H_1} = \left( \frac{P_0}{P_1} \right) \left(\frac{\mathcal{M}_1}{\mathcal{M}_0} \right)
\left( \frac{\Gamma_1}{\Gamma_0}\right)^2  \, .
\end{equation}
Note that this relation does not account for mode visibilities. These will be considered in Section~\ref{Method}.

Equation~(\ref{rapport_hauteurs}) deserves some comments. Firstly, it links the observables $H$ and $\Gamma$ with the quantities $P$ and $\mathcal{M}$, which are only derived from modelling. Secondly, because we require the frequencies $\nu_0$ and $\nu_1$ to be close to each other (say $\vert \nu_0-\nu_1 \vert < \Delta\nu/2$), the shapes of the eigenfunctions in the uppermost layers are very similar. As shown by \cite{Dupret2009}, this implies that the driving are the same:
\begin{equation}
\label{egalite_travail}
P_0 \mathcal{M}_0 \simeq P_1 \mathcal{M}_1 \,. 
\end{equation}
Equation~\ref{egalite_travail} means that, at similar frequencies and with similar mode shapes in the super-adiabatic layers, the work done by the driving source on the modes is the same.  Theoretical computations on evolved stars spanning from the early-subgiant phase to the top of the red giant branch show that this approximation is accurate to better than one percent in the observed frequency range (near $\nu_{max}$) but fails at very low frequencies, where mode amplitudes are too small to be observed. Nevertheless, a thorough investigation of the accuracy of Equation~\ref{egalite_travail} is desirable.

Using Equation~\ref{def:Mode_mass_inertia} and the fact that $\left| \vec \xi_{\ell=0} (R) \right| \simeq \left| \vec \xi_{\ell=1} (R) \right|$, we can consider that mode mass is proportional to mode inertia. Equation~\ref{egalite_travail} can be then be recast as
\begin{equation}
\label{rapport_hauteurs_final_theorique}
% \left(\frac{\mathcal{M}_1}{\mathcal{M}_0} \right)^2 \simeq \frac{H_0}{H_1} \left( \frac{\eta_0}{\eta_1}\right)^2  \, .
 \frac{\mathcal{M}_1}{\mathcal{M}_0} \simeq \frac{I_1}{I_0} \simeq \sqrt{\frac{H_0}{H_1}} \frac{\Gamma_0}{\Gamma_1}  \, .
\end{equation}
This shows that one can derive the ratio of mode inertias (or mode masses) between a radial mode and a neighboring dipole mixed-mode from the observed mode heights and linewidths. 
Using Equation~\ref{def:amplitude}, the inertia ratio can also be rewritten in terms of mode amplitudes and linewidths,
%Nevertheless, anticipating on the following we rewrite Equation~\ref{rapport_hauteurs_final_theorique} using mode amplitude rather than mode heights. Using Equation~(\ref{def:amplitude}), it gives
\begin{equation}
\label{rapport_final}
\frac{I_1}{I_0} \simeq \frac{A_0}{A_1} \sqrt{\frac{\Gamma_0}{\Gamma_1}} \, .
\end{equation}

Note that observed height and linewidth are anti-correlated \citep[e.g.][]{Baudin2005}, which makes it difficult to reliably measure inertia using  Equation~\ref{rapport_hauteurs_final_theorique} when using an algorithm for which the result could be sensitive to the parametrisation (such as MLE). On the other hand, amplitudes and linewidths are weakly correlated so that using Equation~\ref{rapport_final} will provide more reliable results than Equation~\ref{rapport_hauteurs_final_theorique}. %Finally, we note that observed mode amplitudes are intrinsic quantities that must be corrected for visibilities effects, such as described in Sec.~\ref{Method}.

%Nevertheless, anticipating on the following we rewrite Equation~\ref{rapport_hauteurs_final_theorique} using mode amplitude rather than mode heights. Using Equation~(\ref{def:amplitude}), it gives

\section{Observational determination of the mode inertia ratios} 
\label{sec:3}

\begin{table*}
	\caption{Main parameters for KIC 6442183 and KIC 4351319. Model A is for KIC 6442183. Model B is for KIC 4351319 and uses the same physics as model A. References: (a) \cite{Molenda2013arxiv}, (b) \cite{DiMauro2011}, (c) \cite{Benomar2013a}, (d) This study}
	\begin{center}
	\begin{tabular*}{10cm}{cc|c|c|c}
                           &  KIC 6442183                       & Model $A^{(d)}$    & KIC 4351319 &  Model $B^{(b)}$ \\ \hline
$T_{\rm eff} (\rm K)$      &  $5738 \pm 62^{(a)}$ &  $5632$  & $4700 \pm 50^{(b)}$  & $4752$  \\  
$\log\,g$                   &  $4.14 \pm 0.10^{(a)}$            &     $4.01$                     & $3.30 \pm 0.10^{(b)}$      & $3.5$  \\
$\mbox{[Fe/H]}$                   &  $-0.120 \pm 0.050^{(a)}$   &     $-0.036$                           & $0.23 \pm 0.15^{(b)}$    & $0.20$ \\
$\Delta\nu$ $(\mu \rm Hz)$ &  $65.07 \pm 0.09^{(c)}$ &   $65.21$      & $24.38 \pm 0.1^{(d)}$        & $24.6$ \\
$\nu_{\mathrm{max}}$ $(\mu \rm Hz)$ &  $1160 \pm 4^{(c)}$       &                              & $380.7 \pm 4^{(d)}$              &  \\
$\Delta\Pi_1$ $(\rm sec)$  &  $325 \pm 18^{(c)}$                &        $359$             & $97.2^{(d)}$                     & $97.8$  \\
$M$ ($\mbox{M}_\odot$)            &  $\sim 0.94^{(c)}$          &       $1.02$                     & $\sim 1.26^{(d)}$                 & $1.32$  \\
$R$ ($\mbox{R}_\odot$)            &  $\sim 1.60^{(c)}$          &       $1.65$                     & $\sim 3.40^{(d)}$                & $3.39$  \\
$V^2_{\ell=1}$             &  $1.52$  &    & $1.55$  &  \\  \hline              
\label{tab:parameters:Stellar}
\end{tabular*}
\end{center}
\end{table*}

\subsection{Method}
\label{Method}

Here, we show how to infer mode-inertia ratios from \emph{Kepler} observations. The power spectra were prepared using methods described by \cite{Garcia2011}. 
We will first consider the star KIC 6442183, which was analyzed by \cite{Benomar2013a} using Quarters 5 to 7 (nine months). It was selected due to its high signal-to-noise oscillation spectrum and because it does not show large rotational splittings that would result in a more complex power spectrum. Its global seismic characteristics are given in Table~\ref{tab:parameters:Stellar}, and show KIC 6442183 to be an early subgiant star of about one solar mass. 

%\textbf{(known as `Pooh' within the Kepler Asteroseismic Science Consortium)}
We also considered KIC 4351319, a red giant star already analyzed by \cite{DiMauro2011} using Q3.1 data (one month). For this work, we re-analyzed the power spectrum using Q3 to Q14 data (three years) and adopting a Markov Chain Monte-Carlo sampler \cite[e.g.][]{Benomar2009,Handberg2011}, allowing precise measurements of all mode parameters. Table~\ref{tab:parameters:Stellar} summarizes the global characteristics of this star.

%Note that the measured amplitudes are not the bolometric ones and must therefore be corrected by taking the \emph{Kepler} spectral response into account. 
Note that the measured amplitudes are not bolometric and must therefore be corrected for the \emph{Kepler} spectral response. Moreover, geometric and limb darkening effects must be considered. Consequently, Equation~\ref{rapport_final} is modified to become
\begin{equation} 
\label{eq:inertia_ratio:corr}
	\frac{I_1}{I_0} \simeq V_1 \, \frac{A_0}{A_1} \, \sqrt{\frac{\Gamma_0}{\Gamma_1}} \, , 
\end{equation}
where the coefficient $V_1$ (hereafter called visibility)  is defined as  $V^2_{1}=H^{th}_{1}/H^{th}_{0}$ with $H^{th}_{0}$ and $H^{th}_{1}$ being theoretical heights for $\ell=0$ and $\ell=1$ modes. 
To compute $V_1$, we use the results of \cite{Ballot2011}, which considered the spectral response of the \emph{Kepler} spacecraft and include limb darkening and geometric effects. \cite{Ballot2011}, based on the work of \cite{Michel2009}, computed a grid of values for $V_1$ as a function of $T_{\rm eff}$, $\log\,g$ and [Fe/H]. By interpolating this grid, we obtain the visibilities for KIC 6442183 and KIC 4351319 reported in Table \ref{tab:parameters:Stellar} \citep[see ][for a discussion about observed visibilities]{Mosser2012b}.

Using Equation~\ref{eq:inertia_ratio:corr}, together with the mode linewidths and amplitudes measured from \emph{Kepler} observations, it is possible to derive the mode-inertia ratios. However, one must also consider how to compute the ratio of amplitudes $A_0/A_1$ and of linewidths $\Gamma_0/\Gamma_1$. To determine these ratios, we first interpolated $A_0(\nu_0)$ and $\Gamma_0(\nu_0)$ at $\nu_1$ between consecutive radial modes. 
This relies on the fact that amplitudes and linewidths of radial modes vary smoothly from mode to mode \citep[e.g.][]{JCD2004,Dupret2009}.

To robustly estimate the uncertainties, the calculation of the mode-inertia ratio involved the probability density functions of the parameters from the MCMC process. The median of the probability density function defined the most likely solution, while confidence intervals at $\pm \sigma$ were computed using the cumulative distribution function and used as the lower and upper uncertainties.

\subsection{Results}
\label{Results}

\begin{figure*}[]
\begin{center}
\includegraphics[angle=90,totalheight=6.5cm]{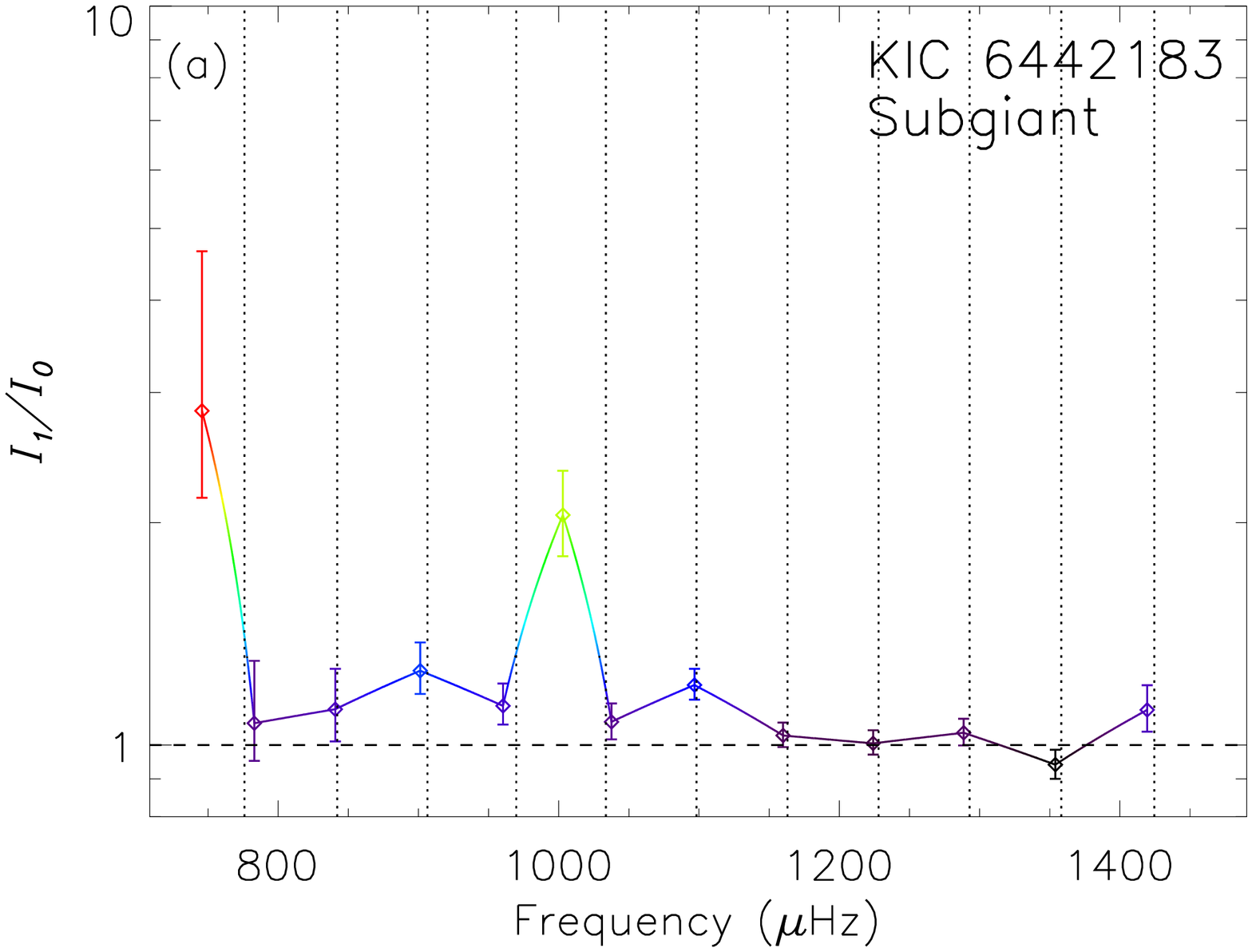}
\includegraphics[angle=90,totalheight=6.5cm]{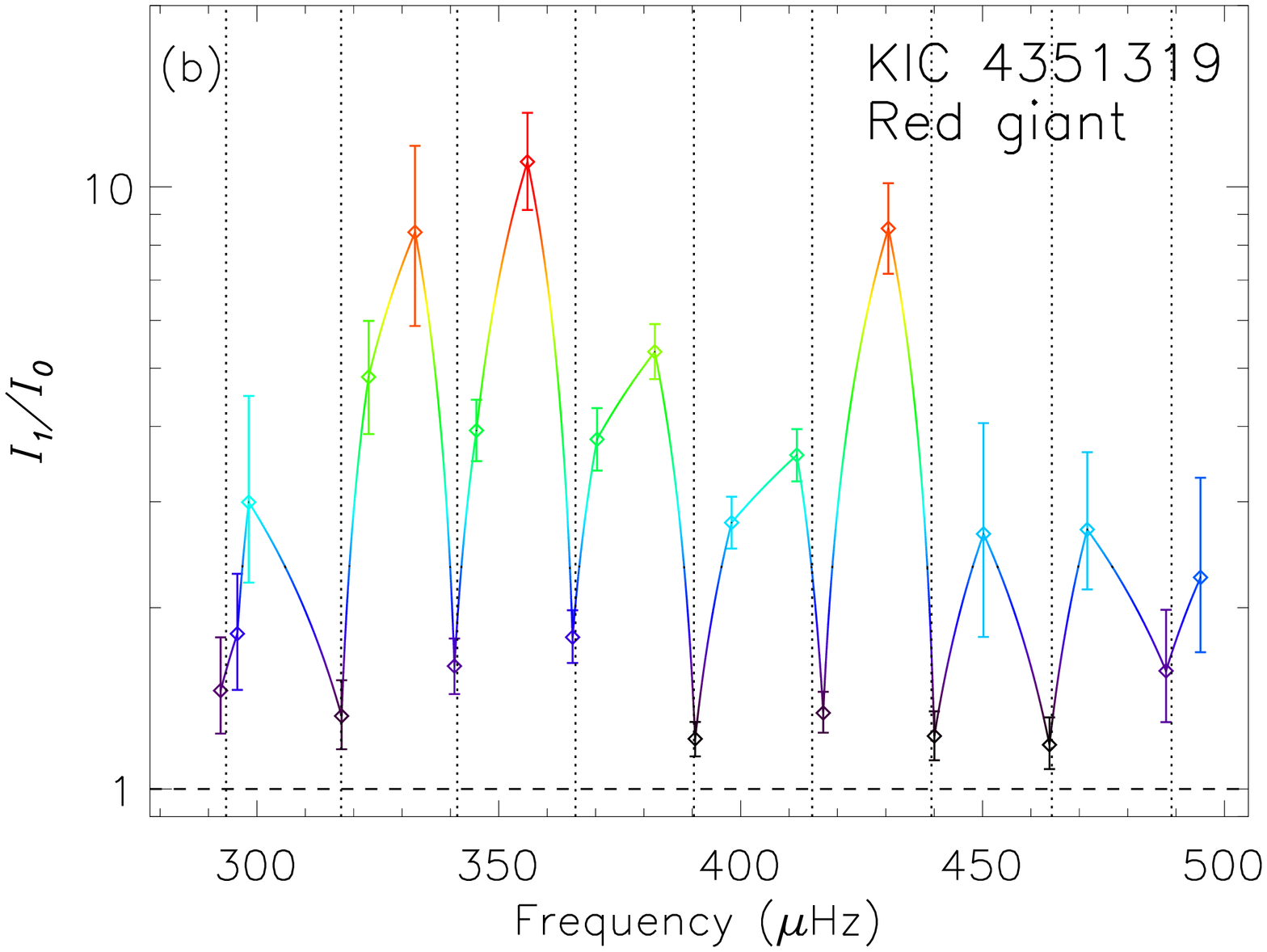}
\includegraphics[angle=90,totalheight=6.5cm]{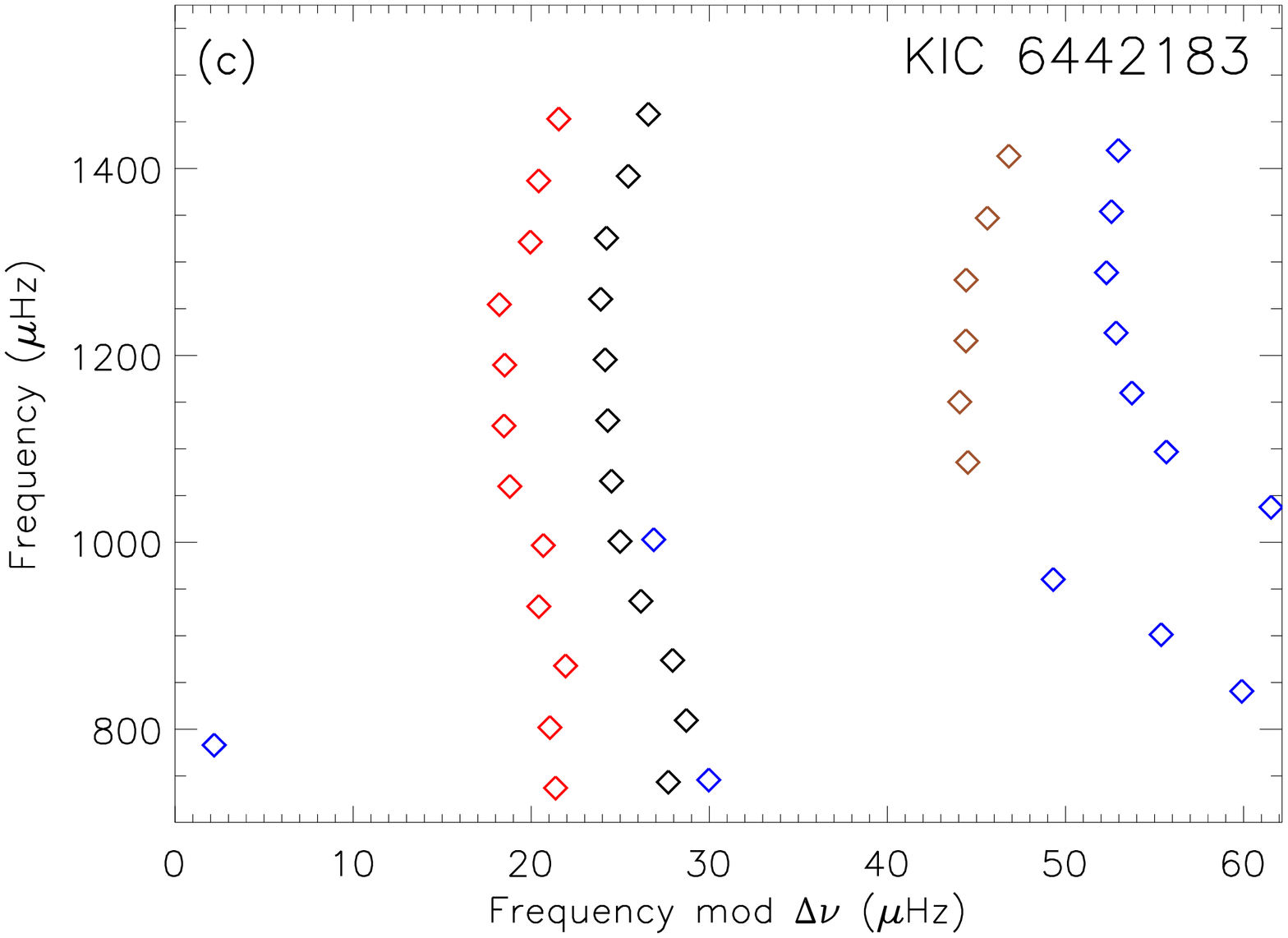}
\includegraphics[angle=90,totalheight=6.5cm]{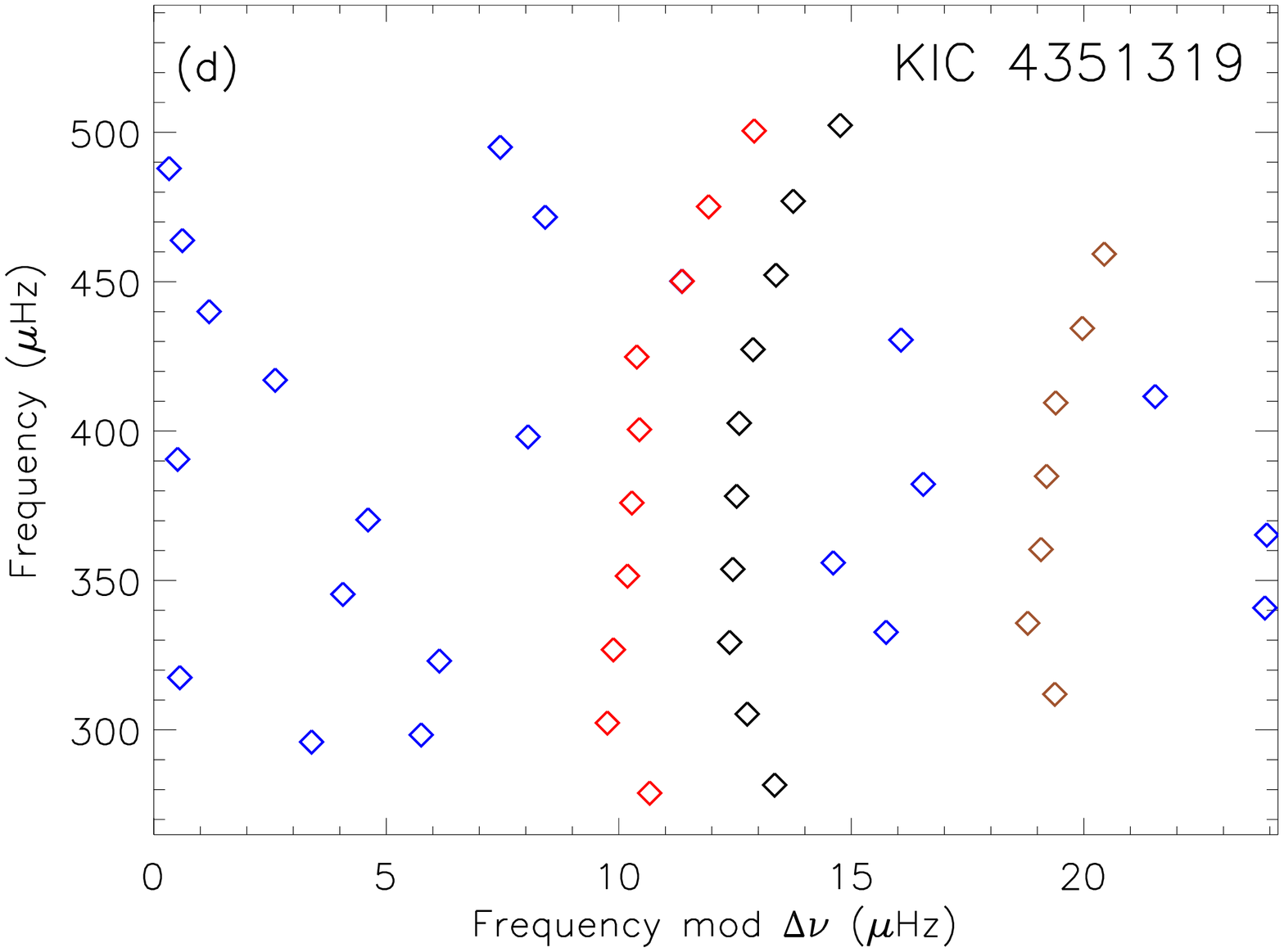}
\end{center}
%\caption{Ratio of the mode inertia between $\ell=1$ and $\ell=0$ measured for KIC 6442183 (a)  and KIC 4351319 (b) as a function of frequency, as well as \'echelle diagrams of observed frequencies for these stars (c, d), with $\ell=0, 1,2,3$ in black, blue, red, brown diamonds respectively. In inertia plots, approximate frequencies of the pure $\ell=1$ p modes ($\nu_{p,1}$) are indicated with vertical dashed lines. Warmer colors indicate high ratio, cold colors indicate lower ratio.}
\caption{Inertia ratio between $\ell=1$ and $\ell=0$ modes measured for KIC 6442183 (a) and KIC 4351319 (b) against frequency.  Approximate positions of the pure $\ell=1$ p modes ($\nu_{p,1}$) are indicated by vertical dashed lines. \'Echelle diagrams of observed frequencies for KIC 6442183 (c)  and KIC 4351319 (d) with $\ell=0, 1,2,3$ modes reported in black, blue, red, brown diamonds respectively.  Warmer colors indicate high ratio, cold colors indicate lower ratio.} 
\label{fig:Inertia_obs}
\end{figure*}

In Fig.~\ref{fig:Inertia_obs} we show the observed mode-inertia ratios for KIC 6442183 and KIC 4351319 as a function of frequency, as well as their \'echelle diagrams. Vertical lines are half-way between consecutive $\ell=0$ p-mode frequencies ($\nu_{p,0}$) and so give good estimates for the $\ell=1$ pure p-mode frequencies ($\nu_{p,1}$).% \textbf{Here Dennis suggests to say more about $\nu_{p,1}$}. 

For the subgiant, the mixed modes closest to the $\ell=1$ p-mode frequencies have the lowest mode-inertia ratios ($I_1/I_0$ remains close to unity), while for mixed modes that depart significantly from the p modes, $I_1/I_0$ increases up to about two. 

In the red giant, the density of g modes is greater than the density of the p modes, allowing us to better trace variations of the mode inertia. $I_1/I_0$ ranges between $1.5$ and $10$ because, in this regime, all modes are strongly mixed (none are pure p modes).

These behaviors are due to the difference in evolutionary stage. As a star evolves from subgiant to red giant, the density contrast between the core and the envelop increases, and so do the mode-inertia ratios \citep{Dupret2009}. This is also consistent with theoretical computations of mode inertias in evolved solar-like stars \citep[e.g.][]{Dupret2009,Montalban2010}. 

\section{Comparison with stellar models} \label{sec:4}

In this section, we show how inertias can provide new diagnostics on the internal structures of evolved stars. To this end, we first compare the observed and modeled values of $I_1/I_0$. We then discuss the potential diagnostics on the stars internal structure. 

\subsection{Comparison between observed and modeled mode-inertia ratios}
\label{comparaison}

\begin{figure*}[t]
\begin{center}
    \includegraphics[angle=90,totalheight=10cm]{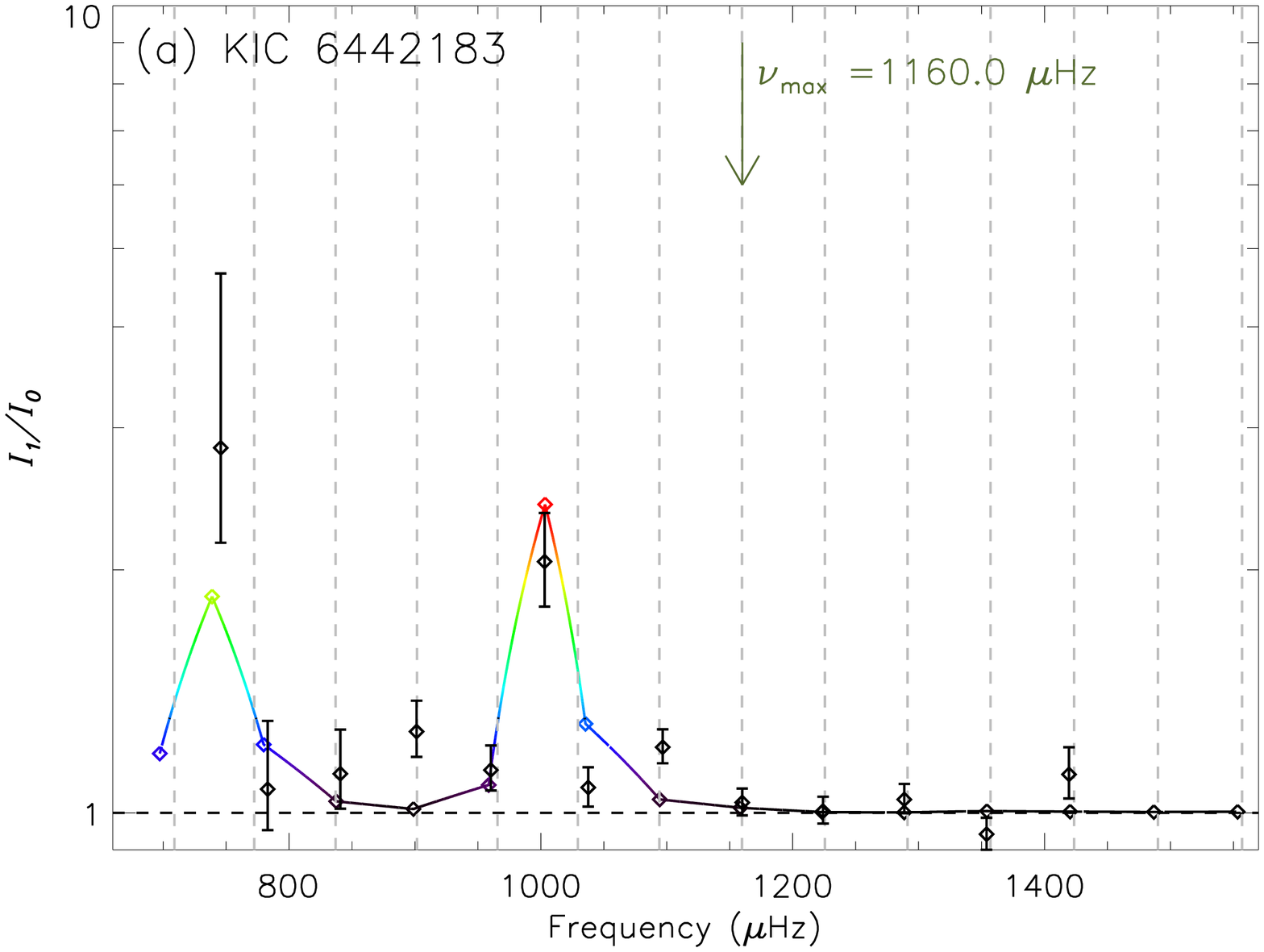} \\
    \includegraphics[angle=90,totalheight=10cm]{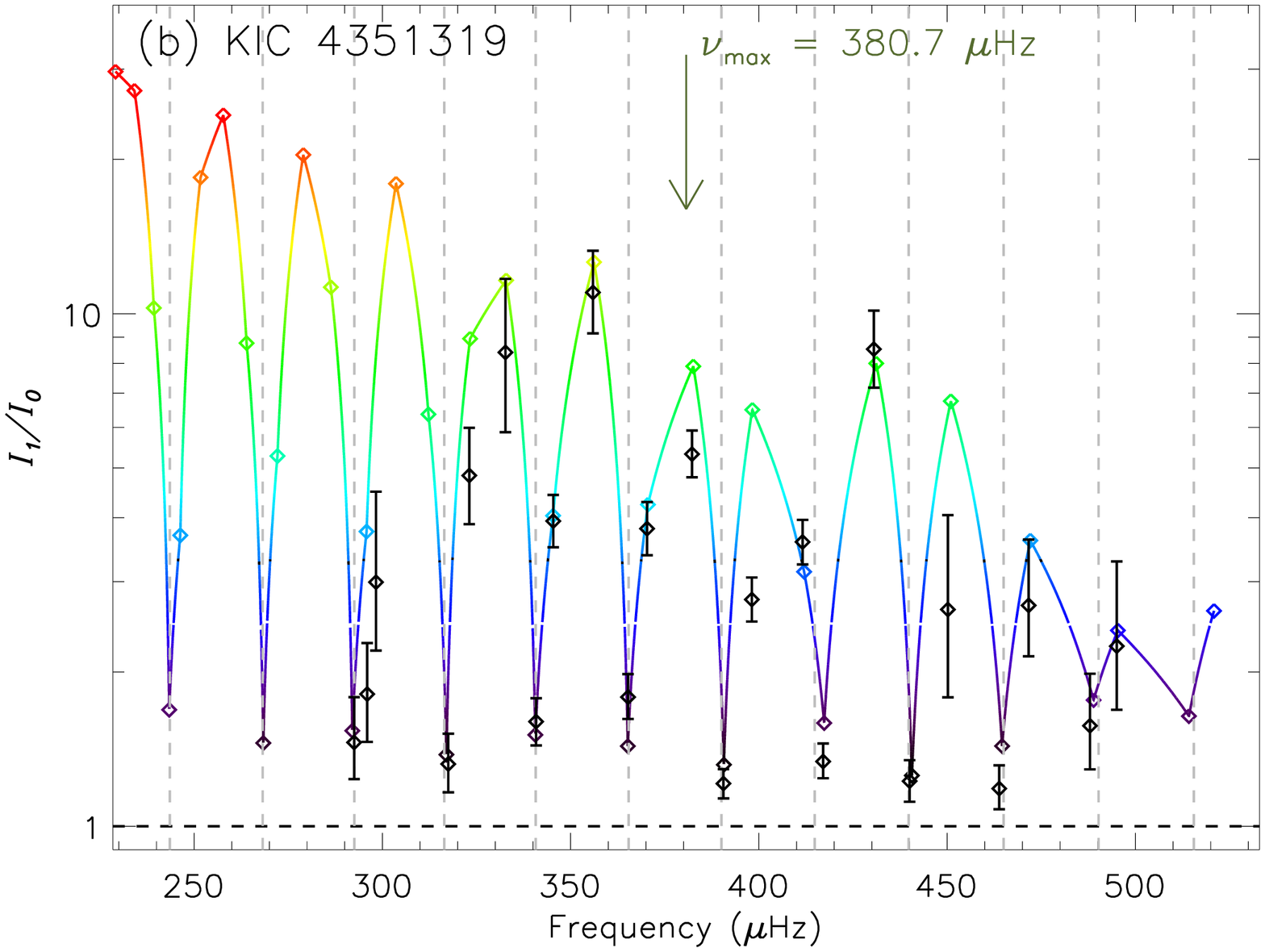}
\end{center}
	%\caption{Measured inertia ratio between $\ell=1$ and $\ell=0$ against frequency (black diamonds) for KIC 4351319 (form \ref{fig:Inertia_obs}). Superimposed, the best model obtained by \cite{DiMauro2011} (colored points connected by lines).}
	\caption{Measured inertia ratio between $\ell=1$ and $\ell=0$ modes against frequency (black diamonds) for KIC 4351319 (a) and KIC 6442183 (b). Theoretical inertia ratio are superimposed (colored points connected by lines). Dashed line indicates the lower limit of inertia ratio (pure p modes). The higher is the inertia, the stronger the mixed character of the mode.}
	\label{fig:Inertia-4351319}
\end{figure*}

%\textbf{KIC 4351319 is a star that was already modeled using seismic constraints by \cite{DiMauro2011}}. Their model was computed using the ASTEC evolution code \citep{JCD2008a} and the associated eigenfrequencies and eigenfunctions were computed using the ADIPLS oscillation code \citep{JCD2008b}. The detailed input physics of the model was described by \cite{DiMauro2011} (see their Table 1 and model 1). 
%The model of KIC 4351319 was selected to reproduce spectroscopic constraints (effective temperature and surface gravity), the set of all the observed frequencies, as well as the large ($\Delta \nu$) and small separations ($\delta_{02}$). 

%\textbf{KIC 6442183 was also modeled using the same physics and method for the purpose of this paper.} The global parameters of the models are reported in Table~\ref{tab:parameters:Stellar}. 

The structure models of KIC 6442183 and KIC 4351319 were computed using the ASTEC evolution code \citep{JCD2008a} and the associated eigenfrequencies and eigenfunctions were computed using ADIPLS oscilation code \citep{JCD2008b}. The input physics used for the modelling is described in detail by \cite{DiMauro2011}. 
The selected models well reproduce spectroscopic constraints (effective temperature and surface gravity), and the set of all the observed frequencies, as well as the large ($\Delta \nu$) and small separations ($\delta_{02}$). For KIC 4351319, we used the evolutionary model previously computed by \cite{DiMauro2011}.
The global parameters of the models are reported in Table~\ref{tab:parameters:Stellar}.

In Fig.~\ref{fig:Inertia-4351319}, the ratio $I_1/I_0$ derived from the modelling is compared with observations and we see very good agreement for most modes. 
This strengthens the validity of our method to derive the ratios $I_1/I_0$ from the observed mode amplitudes and linewidths, because the models were \emph{not} chosen to reproduce the mode inertias but only to match mode frequencies. It also suggests that a more accurate model of these star could be derived using both  frequencies and inertias.
%the model of KIC 4351319 from \cite{DiMauro2011} was \emph{not} intended to reproduce the mode inertias but only to match mode frequencies. It also suggests that a more accurate model of this star could be derived using both inertia and frequencies.

\subsubsection{Small separation and period spacing diagnostics using collapsed \'echelle diagrams}
\label{diagnostic}

\begin{figure*}[]
    \includegraphics[angle=90,totalheight=6.5cm]{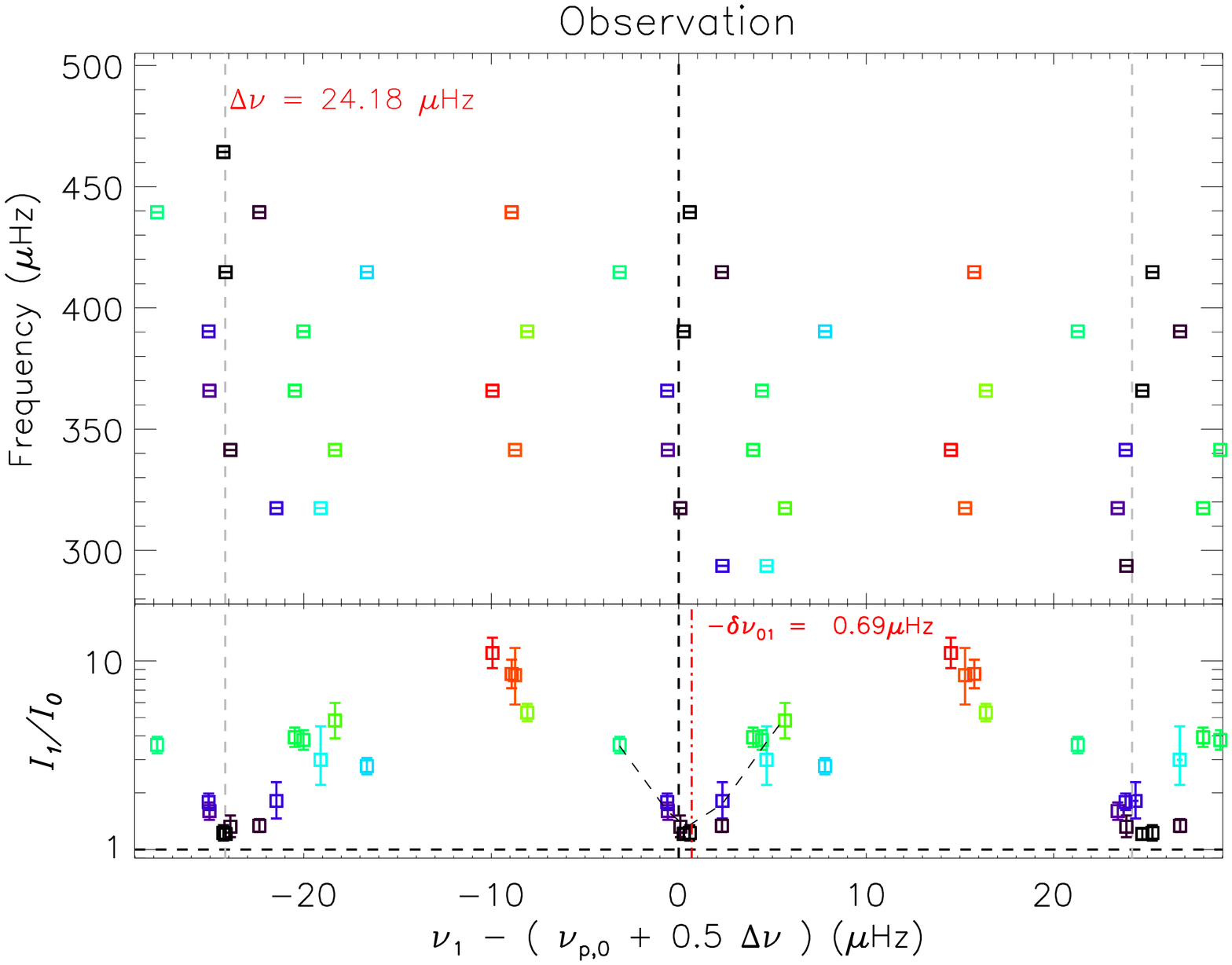}
    \includegraphics[angle=90,totalheight=6.5cm]{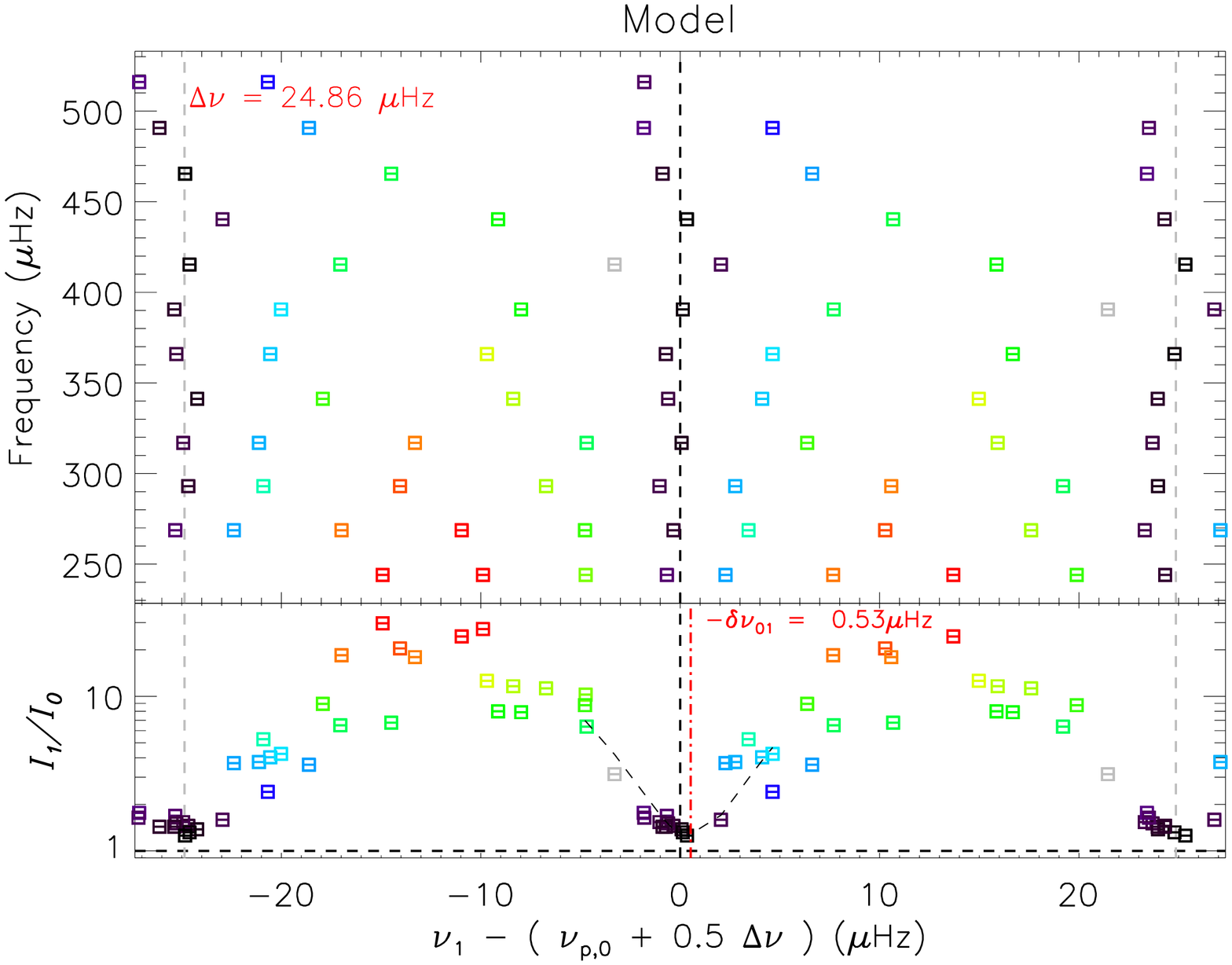}
    \caption{Frequency-\'echelle (top) and collapsed frequency-\'echelle diagram for KIC 4351319 for observations (left) and for model B (right). The minima of inertia are spaced by $\Delta\nu$. The distance to $0$ in the abscissa measures $\delta\nu_{01}$.}
	\label{fig:Frequency-Echelle-4351319}
\end{figure*}

\begin{figure*}[]
    \includegraphics[angle=90,totalheight=6.5cm]{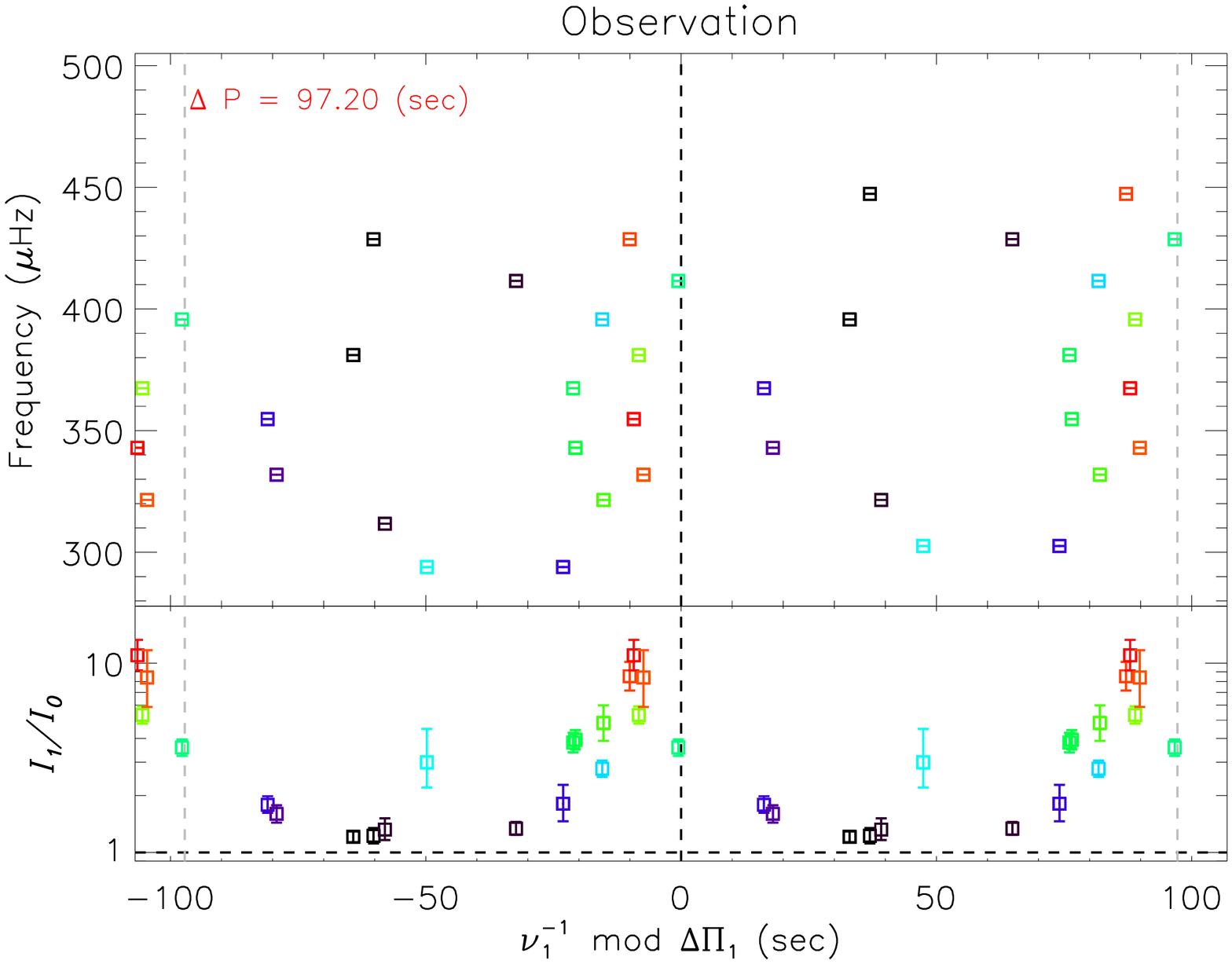}
    \includegraphics[angle=90,totalheight=6.5cm]{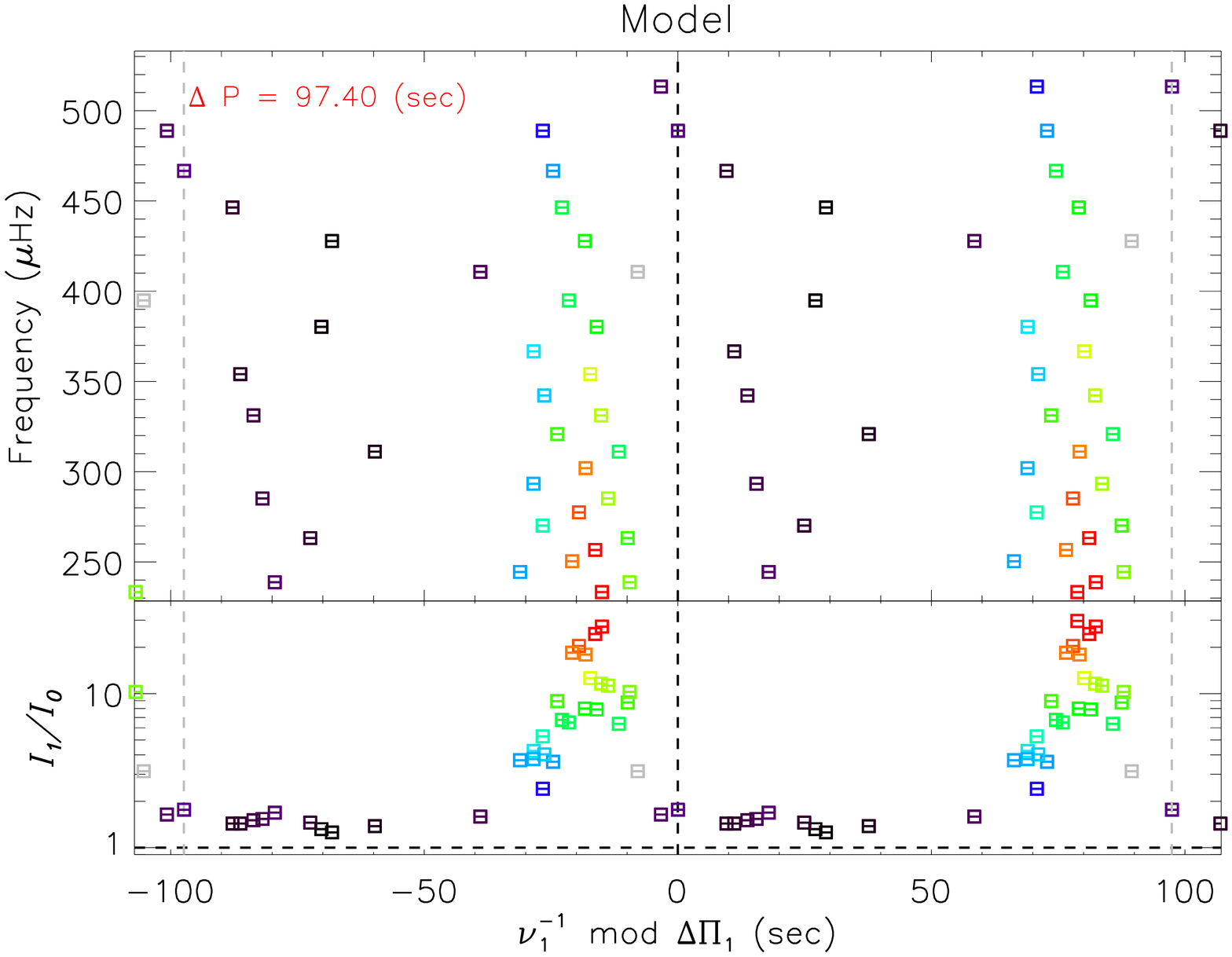}
    \caption{Period-\'echelle (top) and collapsed period-\'echelle diagram for KIC 4351319 for observations (left) and for model B (right). The maxima of inertia are spaced by $\Delta\Pi_1$.}% The distance to $0$ of the maximum of inertia in the x-axis measures $\epsilon_g$.}
	\label{fig:Period-Echelle-4351319}
\end{figure*}

In the following, we focus on the red giant KIC 4351319, since the variations in $I_1 / I_0$ are greater. \'Echelle diagrams, either in frequency or in period %(as well as collapsed \'echelle)
help to visualise these variations and provide diagnostics on mixed modes (see Fig.~\ref{fig:Frequency-Echelle-4351319} and Fig.~\ref{fig:Period-Echelle-4351319}). 

Frequency-\'echelle diagrams are produced by representing the frequencies as a function of the difference $\nu_1 - (\nu_{p,0} + 0.5\Delta\nu)$, $\Delta\nu$ being measured using the radial frequencies. Note that in contrast to a normal \'echelle diagram (Fig.~\ref{fig:Inertia_obs}~c,d), here we remove any curvature in the radial ridge by substracting the measured radial mode frequency at each order. Collapsing the \'echelle diagram vertically and plotting inertia on the ordinate (bottom panels) reveals clearly the variation of the inertia with frequency. 

Based on the asymptotic relation for p mode, it is conventional to define the small separation $\delta\nu_{01} = -\nu_{p, 1} + (\nu_{p,0} + 0.5\Delta\nu)$. As shown in Fig.\ref{fig:Inertia_obs}, $\ell=1$ modes with the lowest inertia behave almost as pure p-modes. Therefore, the frequency position of the inertia minimum allows us to estimate $\delta\nu_{01}$.
%Therefore, the most p-like modes having the lowest inertia, the position of the minimum of inertia in the folded \'echelle diagram allows us to measure the mean value of $\delta\nu_{01}$. 
For KIC 4351319, we measured $\delta\nu_{01}=-0.69 \pm 0.30$ $\mu$Hz by fitting a second order polynomial within the range $[-\Delta\nu/4,  \Delta\nu/4]$ on the collapsed \'echelle. For comparison, the same approach on model frequencies gives $\delta\nu_{01}=-0.53 \pm 0.24$ $\mu$Hz (cf. Fig.\ref{fig:Frequency-Echelle-4351319}). Note that negative values are found in most evolved stars \citep[\emph{e.g.}][]{Bedding2010, Mosser2011b, Corsaro2012}. %\textbf{Negative values of small spacing in red giants have been already reported by \cite{Mosser2012} (\textcolor{red}{En fait je n arrive pas a trouver une reference dans ces deux derniers papiers sur les modes mixtes... tu te souvient ou il a mis ca?)} which strenghthen the validity of our approach}\textcolor{red}{Est ce que l on peux en dire plus a propos des valeurs negatives?}. 
%, a quantity that is sensitive to the core properties. 

%Note also that when $\nu_1 - (\nu_{p,0} + 0.5\Delta\nu) \simeq \Delta\nu/2$, modes are more g-like. The abruptness of the transition between these two extremes depends on the coupling strength and therefore on the size of the evanescent zone between the p- and g-mode cavities.

The period-\'echelle diagram (Fig.\ref{fig:Period-Echelle-4351319}) 
presents the frequency ratios as a function of the period $\nu^{-1}_1$ mod $\Delta\Pi_1$, where $\Delta\Pi_1$ is the average period spacing of the $\ell=1$ g-mode frequencies \citep{Bedding2011Nature} . %$\Delta\Pi_1$ is \emph{a priori} not known, however, because 
The maxima of mode inertias are separated by $\Delta\Pi_1$ in the period-\'echelle diagram, so by collapsing vertically over the correct value of the period-spacing, the mode-inertia-period plot shows a repetitive pattern of peaks, spaced by $\Delta \Pi_1$ (the pattern period). This is therefore a new method to measure the period spacing in evolved stars, with a precision of better than $1\%$. The higher the g mode density, the smaller the uncertainty. For KIC 4351319 we found $\Delta\Pi_1 = 97.2$ sec. 
%This period spacing is compatible with the value derived by other methods such as the asymptotic approach by \cite{Mosser2012a}, with which we found $\Delta\Pi_1 = 96.5$ sec. We also considered period spacing of the models. The value found by integrating the Brunt V\"ais\"al\"a frequency \citep[e.g.][]{JCD2012} is $97.80$ sec, while we found 97.4 sec with our method (cf. Fig.\ref{fig:Period-Echelle-4351319}).
This period spacing is compatible with that derived by other methods such as the asymptotic approach by \cite{Mosser2012a}, which applied to our data gives $\Delta\Pi_1 = 96.5$ sec. We also computed the period spacing of the model 1 of \cite{DiMauro2011} and found a value of $97.4$ sec, which is compatible with the value of 97.8 sec found by integrating the Brunt V\"ais\"al\"a frequency \citep[e.g.][]{JCD2012}. Relative uncertainties are $\simeq 0.1$ \% here.

\section{Discussion and conclusion} \label{sec:5}

In this Letter, we presented a method to measure mode-inertia ratios, a quantity which until now, was only derived from modelling. This diagnostic is based on the detection and precise measurement of mixed-mode amplitudes and linewidths. Using \emph{Kepler} data, we demonstrated that it is possible to measure precisely the mode inertias of subgiants and early red giants. These can be used to determine the period spacing and the coupling between the $p$ and $g$ cavities of mixed modes. Those results were validated by comparing mode-inertia ratios derived from the observations and from stellar and pulsational modelling of KIC 4351319 and KIC 6442183. 

%As shown in Sects.~\ref{comparaison}, there is a good agreement between the observed and model inertia ratios. This results is enhanced by the fact that observed mode inertias allows us to measure the mean small separation and period spacing (see Sect.~\ref{diagnostic}) in agreement with both the model of KIC and the values inferred directly from the measured frequencies. 

To go a step further, and to better evaluate which constraints inertias may provide,  %have more insight into the constraints mode inertia ratios can provide, 
it is useful to adopt an asymptotic approach. For this, we follow the work of \cite{Goupil2013} based on the asymptotic formalism of \cite{Shibahashi1979} for mixed modes. This allows us to write 
\begin{equation}
\label{rapport_asymptotique}
\frac{I_1}{I_0} \approx   
\left\{1  + \frac{1}{4} q^2 \left(\frac{\kappa_g}{\kappa_p}\right) \left(\frac{\cos \kappa_p}{\cos \kappa_g }\right)^2 \right\},
\end{equation}
where $q$ measures the coupling between the $p$- and $g$-mode cavities, as introduced by \cite{Mosser2012a}, and 
\begin{eqnarray} \label{def_tp_tg}
\kappa_p \approx \frac{\pi \; \nu}{\Delta \nu} \, , \quad {\rm and} \quad
\kappa_g  \approx \frac{\pi}{\Delta \Pi_1 \; \nu} - \frac{\pi}{2}\, ,
\end{eqnarray}
%where $\sigma=2\pi\nu$ is the angular frequency. 
where $\nu$ is the frequency. In Equation~\ref{rapport_asymptotique} and Equation~\ref{def_tp_tg}, the period of the oscillation in the ratio $I_1 / I_0$ is given by $\kappa_g$ (thus by the period spacing $\Delta \Pi_1$) appearing as the argument of the cosine function in Equation~\ref{rapport_asymptotique}. This is due to the fact that the term $\cos (\kappa_p)$ is almost constant over the considered frequency range.  %
 
Moreover, the minimum value of the inertia ratio is determined by the ratio $\kappa_g /\kappa_p$ in Equation~\ref{rapport_asymptotique}, which depends on the inverse of the squared frequency ($1/\nu^2$). We also note that the right-hand side of this equation is directly proportional to $q^2$, which is related to the strength of the coupling and characterizes the evanescent region.  Therefore, Equation~\ref{rapport_asymptotique} and Equation~\ref{def_tp_tg} confim that the ratio $I_1 / I_0$ can be used as a tool to derive the period spacing and also to constrain the evanescent region between the $p$ and $g$ cavities. This is in agreement with the results of Sect.~\ref{diagnostic}.

The ratio $I_1 / I_0$ provides an alternative  to using mixed mode frequencies and gives essentially the same constraints \cite[see Equation~9 of][]{Mosser2012a}. However, we note that the relationship of $I_1 / I_0$ as a function of frequency to (i) the period spacing ($\Delta \Pi_1$) and (ii) the properties of the inner-most layers characterised by $q$ is different than Equation~9 of \cite{Mosser2012a}. Therefore, the two observational constraints are complementary, potentially providing stronger constraints on the stellar structure. Further investigations are required to evaluate the advantages of observed inertia as constrains on the stellar structure.

%It is well known that frequencies depend highly on poorly modeled surface effects \cite[e.g.][]{JCD1997,Rosenthal1999}. For the ratio $I_1 / I_0$, such a dependence is not yet clear, but a combination of mode inertia ratios of mode frequencies is potentially a way to better understand those surface effects. 
We stress that inertia ratios are sensitive to non-adiabacitic effects, not accounted by \cite{Ballot2011} when computing visibilities \citep{Toutain1993}. \cite{Mosser2012b} noted that height ratios can differ by $10\%$ from adiabatically computed values, possibly biasing inertia ratios by $5\%$.

Our approach requires well-resolved mixed modes and well-identifed mode rotationnal splittings, which can only be achieved with long observations. This limitation is severe for the more evolved stars because their modes have the longest lifetimes.  
The observations used here, spanning more than three years, resolve mixed modes down to $\nu_{\mathrm{max}} \simeq 120$ $\mu$Hz. We expect to push this limit even further with the latest \emph{Kepler} observations (reaching four years).% We therefore remain confident that with the exploitation of the full \emph{Kepler} observations, such diagnostic can be extended along the red giant branch.

\vspace{0.5cm}

{ \it Acknowledgments. "KB, BM, MJG, and RS acknowledge the ANR (Agence Nationale de la
Recherche, France) program IDEE (n$^\circ$ ANR-12-BS05-0008) ``Interaction Des
\'Etoiles et des Exoplan\`etes'' as well as financial support from
"Programme National de Physique Stellaire" (PNPS) of CNRS/INSU, France".
 RAG acknowledge the support of the European Community's Seventh Framework Programme (FP7/2007-2013) under grant agreement no. 269194 (IRSES/ASK).
 }

\bibliographystyle{apj}
%\bibliography{mybib_Inertia}

\begin{thebibliography}{0}
\expandafter\ifx\csname natexlab\endcsname\relax\def\natexlab#1{#1}\fi

\end{thebibliography}


\begin{thebibliography}{47}
\expandafter\ifx\csname natexlab\endcsname\relax\def\natexlab#1{#1}\fi

\bibitem[{{Aizenman} {et~al.}(1977){Aizenman}, {Smeyers}, \&
  {Weigert}}]{Aizenman1977}
{Aizenman}, M., {Smeyers}, P., \& {Weigert}, A. 1977, \aap, 58, 41

\bibitem[{{Appourchaux} {et~al.}(2008){Appourchaux}, {Michel}, {Auvergne},
  {Baglin}, {Toutain}, {Baudin}, {Benomar}, {Chaplin}, {Deheuvels}, {Samadi},
  {Verner}, {Boumier}, {Garc{\'{\i}}a}, {Mosser}, {Hulot}, {Ballot}, {Barban},
  {Elsworth}, {Jim{\'e}nez-Reyes}, {Kjeldsen}, {R{\'e}gulo}, \&
  {Roxburgh}}]{Appourchaux2008}
{Appourchaux}, T., {et~al.} 2008, \aap, 488, 705

\bibitem[{{Appourchaux} {et~al.}(2012){Appourchaux}, {Chaplin},
  {Garc{\'{\i}}a}, {Gruberbauer}, {Verner}, {Antia}, {Benomar}, {Campante},
  {Davies}, {Deheuvels}, {Handberg}, {Hekker}, {Howe}, {R{\'e}gulo},
  {Salabert}, {Bedding}, {White}, {Ballot}, {Mathur}, {Silva Aguirre},
  {Elsworth}, {Basu}, {Gilliland}, {Christensen-Dalsgaard}, {Kjeldsen},
  {Uddin}, {Stumpe}, \& {Barclay}}]{Appourchaux2012}
---. 2012, \aap, 543, A54

\bibitem[{{Baglin} {et~al.}(2006{\natexlab{a}}){Baglin}, {Auvergne}, {Barge},
  {Deleuil}, {Catala}, {Michel}, {Weiss}, \& {COROT Team}}]{Baglin2006a}
{Baglin}, A., {Auvergne}, M., {Barge}, P., {Deleuil}, M., {Catala}, C.,
  {Michel}, E., {Weiss}, W., \& {COROT Team}. 2006{\natexlab{a}}, in ESA
  Special Publication, Vol. 1306, ESA Special Publication, ed. M.~{Fridlund},
  A.~{Baglin}, J.~{Lochard}, \& L.~{Conroy}, 33

\bibitem[{{Baglin} {et~al.}(2006{\natexlab{b}}){Baglin}, {Auvergne},
  {Boisnard}, {Lam-Trong}, {Barge}, {Catala}, {Deleuil}, {Michel}, \&
  {Weiss}}]{Baglin2006b}
{Baglin}, A., {et~al.} 2006{\natexlab{b}}, in COSPAR Meeting, Vol.~36, 36th
  COSPAR Scientific Assembly, 3749

\bibitem[{{Ballot} {et~al.}(2011){Ballot}, {Barban}, \& {van't
  Veer-Menneret}}]{Ballot2011}
{Ballot}, J., {Barban}, C., \& {van't Veer-Menneret}, C. 2011, \aap, 531, A124

\bibitem[{{Baudin} {et~al.}(2005){Baudin}, {Samadi}, {Goupil}, {Appourchaux},
  {Barban}, {Boumier}, {Chaplin}, \& {Gouttebroze}}]{Baudin2005}
{Baudin}, F., {Samadi}, R., {Goupil}, M.-J., {Appourchaux}, T., {Barban}, C.,
  {Boumier}, P., {Chaplin}, W.~J., \& {Gouttebroze}, P. 2005, \aap, 433, 349

\bibitem[{{Beck} {et~al.}(2011){Beck}, {Bedding}, {Mosser}, {Stello}, {Garcia},
  {Kallinger}, {Hekker}, {Elsworth}, {Frandsen}, {Carrier}, {De Ridder},
  {Aerts}, {White}, {Huber}, {Dupret}, {Montalb{\'a}n}, {Miglio}, {Noels},
  {Chaplin}, {Kjeldsen}, {Christensen-Dalsgaard}, {Gilliland}, {Brown},
  {Kawaler}, {Mathur}, \& {Jenkins}}]{Beck2011Science}
{Beck}, P.~G., {et~al.} 2011, Science, 332, 205

\bibitem[{{Bedding} {et~al.}(2010){Bedding}, {Huber}, {Stello}, {Elsworth},
  {Hekker}, {Kallinger}, {Mathur}, {Mosser}, {Preston}, {Ballot}, {Barban},
  {Broomhall}, {Buzasi}, {Chaplin}, {Garc{\'{\i}}a}, {Gruberbauer}, {Hale}, {De
  Ridder}, {Frandsen}, {Borucki}, {Brown}, {Christensen-Dalsgaard},
  {Gilliland}, {Jenkins}, {Kjeldsen}, {Koch}, {Belkacem}, {Bildsten}, {Bruntt},
  {Campante}, {Deheuvels}, {Derekas}, {Dupret}, {Goupil}, {Hatzes}, {Houdek},
  {Ireland}, {Jiang}, {Karoff}, {Kiss}, {Lebreton}, {Miglio}, {Montalb{\'a}n},
  {Noels}, {Roxburgh}, {Sangaralingam}, {Stevens}, {Suran}, {Tarrant}, \&
  {Weiss}}]{Bedding2010}
{Bedding}, T.~R., {et~al.} 2010, \apjl, 713, L176

\bibitem[{{Bedding} {et~al.}(2011){Bedding}, {Mosser}, {Huber},
  {Montalb{\'a}n}, {Beck}, {Christensen-Dalsgaard}, {Elsworth},
  {Garc{\'{\i}}a}, {Miglio}, {Stello}, {White}, {De Ridder}, {Hekker}, {Aerts},
  {Barban}, {Belkacem}, {Broomhall}, {Brown}, {Buzasi}, {Carrier}, {Chaplin},
  {di Mauro}, {Dupret}, {Frandsen}, {Gilliland}, {Goupil}, {Jenkins},
  {Kallinger}, {Kawaler}, {Kjeldsen}, {Mathur}, {Noels}, {Aguirre}, \&
  {Ventura}}]{Bedding2011Nature}
---. 2011, \nat, 471, 608

\bibitem[{{Belkacem} {et~al.}(2009){Belkacem}, {Samadi}, {Goupil}, {Dupret},
  {Brun}, \& {Baudin}}]{Belkacem2009}
{Belkacem}, K., {Samadi}, R., {Goupil}, M.~J., {Dupret}, M.~A., {Brun}, A.~S.,
  \& {Baudin}, F. 2009, \aap, 494, 191

\bibitem[{{Belkacem} {et~al.}(2006){Belkacem}, {Samadi}, {Goupil}, {Kupka}, \&
  {Baudin}}]{Belkacem2006b}
{Belkacem}, K., {Samadi}, R., {Goupil}, M.~J., {Kupka}, F., \& {Baudin}, F.
  2006, \aap, 460, 183

\bibitem[{{Benomar} {et~al.}(2009){Benomar}, {Appourchaux}, \&
  {Baudin}}]{Benomar2009}
{Benomar}, O., {Appourchaux}, T., \& {Baudin}, F. 2009, \aap, 506, 15

\bibitem[{{Benomar} {et~al.}(2013){Benomar}, {Bedding}, {Mosser}, {Stello},
  {Belkacem}, {Garcia}, {White}, {Kuehn}, {Deheuvels}, \&
  {Christensen-Dalsgaard}}]{Benomar2013a}
{Benomar}, O., {et~al.} 2013, \apj, 767, 158

\bibitem[{{Borucki} {et~al.}(2010){Borucki}, {Koch}, {Basri}, {Batalha},
  {Brown}, {Caldwell}, {Caldwell}, {Christensen-Dalsgaard}, {Cochran},
  {DeVore}, {Dunham}, {Dupree}, {Gautier}, {Geary}, {Gilliland}, {Gould},
  {Howell}, {Jenkins}, {Kondo}, {Latham}, {Marcy}, {Meibom}, {Kjeldsen},
  {Lissauer}, {Monet}, {Morrison}, {Sasselov}, {Tarter}, {Boss}, {Brownlee},
  {Owen}, {Buzasi}, {Charbonneau}, {Doyle}, {Fortney}, {Ford}, {Holman},
  {Seager}, {Steffen}, {Welsh}, {Rowe}, {Anderson}, {Buchhave}, {Ciardi},
  {Walkowicz}, {Sherry}, {Horch}, {Isaacson}, {Everett}, {Fischer}, {Torres},
  {Johnson}, {Endl}, {MacQueen}, {Bryson}, {Dotson}, {Haas}, {Kolodziejczak},
  {Van Cleve}, {Chandrasekaran}, {Twicken}, {Quintana}, {Clarke}, {Allen},
  {Li}, {Wu}, {Tenenbaum}, {Verner}, {Bruhweiler}, {Barnes}, \&
  {Prsa}}]{Borucki2010}
{Borucki}, W.~J., {et~al.} 2010, Science, 327, 977

\bibitem[{{Chaplin} {et~al.}(1998){Chaplin}, {Elsworth}, {Isaak}, {Lines},
  {McLeod}, {Miller}, \& {New}}]{Chaplin1998}
{Chaplin}, W.~J., {Elsworth}, Y., {Isaak}, G.~R., {Lines}, R., {McLeod}, C.~P.,
  {Miller}, B.~A., \& {New}, R. 1998, \mnras, 298, L7

\bibitem[{{Chaplin} {et~al.}(2005){Chaplin}, {Houdek}, {Elsworth}, {Gough},
  {Isaak}, \& {New}}]{Chaplin2005}
{Chaplin}, W.~J., {Houdek}, G., {Elsworth}, Y., {Gough}, D.~O., {Isaak}, G.~R.,
  \& {New}, R. 2005, \mnras, 360, 859

\bibitem[{{Christensen-Dalsgaard}(2004)}]{JCD2004}
{Christensen-Dalsgaard}, J. 2004, \solphys, 220, 137

\bibitem[{{Christensen-Dalsgaard}(2008{\natexlab{a}})}]{JCD2008b}
---. 2008{\natexlab{a}}, \apss, 316, 113

\bibitem[{{Christensen-Dalsgaard}(2008{\natexlab{b}})}]{JCD2008a}
---. 2008{\natexlab{b}}, \apss, 316, 13

\bibitem[{{Christensen-Dalsgaard}(2012)}]{JCD2012}
---. 2012, Astronomische Nachrichten, 333, 914

\bibitem[{{Corsaro} {et~al.}(2012){Corsaro}, {Stello}, {Huber}, {Bedding},
  {Bonanno}, {Brogaard}, {Kallinger}, {Benomar}, {White}, {Mosser}, {Basu},
  {Chaplin}, {Christensen-Dalsgaard}, {Elsworth}, {Garc{\'{\i}}a}, {Hekker},
  {Kjeldsen}, {Mathur}, {Meibom}, {Hall}, {Ibrahim}, \& {Klaus}}]{Corsaro2012}
{Corsaro}, E., {et~al.} 2012, \apj, 757, 190

\bibitem[{{Di Mauro} {et~al.}(2011){Di Mauro}, {Cardini}, {Catanzaro},
  {Ventura}, {Barban}, {Bedding}, {Christensen-Dalsgaard}, {De Ridder},
  {Hekker}, {Huber}, {Kallinger}, {Miglio}, {Montalban}, {Mosser}, {Stello},
  {Uytterhoeven}, {Kinemuchi}, {Kjeldsen}, {Mullally}, \&
  {Still}}]{DiMauro2011}
{Di Mauro}, M.~P., {et~al.} 2011, \mnras, 415, 3783

\bibitem[{{Dupret} {et~al.}(2009){Dupret}, {Belkacem}, {Samadi}, {Montalban},
  {Moreira}, {Miglio}, {Godart}, {Ventura}, {Ludwig}, {Grigahc{\`e}ne},
  {Goupil}, {Noels}, \& {Caffau}}]{Dupret2009}
{Dupret}, M.-A., {et~al.} 2009, \aap, 506, 57

\bibitem[{{Dziembowski}(1971)}]{Dziembowski1971}
{Dziembowski}, W.~A. 1971, \actaa, 21, 289

\bibitem[{{Dziembowski} {et~al.}(2001){Dziembowski}, {Gough}, {Houdek}, \&
  {Sienkiewicz}}]{Dziembowski2001}
{Dziembowski}, W.~A., {Gough}, D.~O., {Houdek}, G., \& {Sienkiewicz}, R. 2001,
  \mnras, 328, 601

\bibitem[{{Garc{\'{\i}}a} {et~al.}(2011){Garc{\'{\i}}a}, {Hekker}, {Stello},
  {Guti{\'e}rrez-Soto}, {Handberg}, {Huber}, {Karoff}, {Uytterhoeven},
  {Appourchaux}, {Chaplin}, {Elsworth}, {Mathur}, {Ballot},
  {Christensen-Dalsgaard}, {Gilliland}, {Houdek}, {Jenkins}, {Kjeldsen},
  {McCauliff}, {Metcalfe}, {Middour}, {Molenda-Zakowicz}, {Monteiro}, {Smith},
  \& {Thompson}}]{Garcia2011}
{Garc{\'{\i}}a}, R.~A., {et~al.} 2011, \mnras, 414, L6

\bibitem[{{Gizon} {et~al.}(2013){Gizon}, {Ballot}, {Michel}, {Stahn},
  {Vauclair}, {Bruntt}, {Quirion}, {Benomar}, {Vauclair}, {Appourchaux},
  {Auvergne}, {Baglin}, {Barban}, {Baudin}, {Bazot}, {Campante}, {Catala},
  {Chaplin}, {Creevey}, {Deheuvels}, {Dolez}, {Elsworth}, {Garcia}, {Gaulme},
  {Mathis}, {Mathur}, {Mosser}, {Regulo}, {Roxburgh}, {Salabert}, {Samadi},
  {Sato}, {Verner}, {Hanasoge}, \& {Sreenivasan}}]{Gizon2013}
{Gizon}, L., {et~al.} 2013, Proceedings of the National Academy of Science,
  110, 13267

\bibitem[{{Goldreich} {et~al.}(1994){Goldreich}, {Murray}, \& {Kumar}}]{GMK94}
{Goldreich}, P., {Murray}, N., \& {Kumar}, P. 1994, \apj, 424, 466

\bibitem[{{Goupil} {et~al.}(2013){Goupil}, {Mosser}, {Marques}, {Ouazzani},
  {Belkacem}, {Lebreton}, \& {Samadi}}]{Goupil2013}
{Goupil}, M.~J., {Mosser}, B., {Marques}, J.~P., {Ouazzani}, R.~M., {Belkacem},
  K., {Lebreton}, Y., \& {Samadi}, R. 2013, \aap, 549, A75

\bibitem[{{Handberg} \& {Campante}(2011)}]{Handberg2011}
{Handberg}, R., \& {Campante}, T.~L. 2011, \aap, 527, A56

\bibitem[{{Libbrecht}(1988)}]{Libbrecht1988}
{Libbrecht}, K.~G. 1988, \apj, 334, 510

\bibitem[{{Metcalfe} {et~al.}(2012){Metcalfe}, {Chaplin}, {Appourchaux},
  {Garc{\'{\i}}a}, {Basu}, {Brand{\~a}o}, {Creevey}, {Deheuvels}, {Do{\v g}an},
  {Eggenberger}, {Karoff}, {Miglio}, {Stello}, {Y{\i}ld{\i}z}, {{\c C}elik},
  {Antia}, {Benomar}, {Howe}, {R{\'e}gulo}, {Salabert}, {Stahn}, {Bedding},
  {Davies}, {Elsworth}, {Gizon}, {Hekker}, {Mathur}, {Mosser}, {Bryson},
  {Still}, {Christensen-Dalsgaard}, {Gilliland}, {Kawaler}, {Kjeldsen},
  {Ibrahim}, {Klaus}, \& {Li}}]{Metcalfe2012}
{Metcalfe}, T.~S., {et~al.} 2012, \apjl, 748, L10

\bibitem[{{Michel} {et~al.}(2009){Michel}, {Samadi}, {Baudin}, {Barban},
  {Appourchaux}, \& {Auvergne}}]{Michel2009}
{Michel}, E., {Samadi}, R., {Baudin}, F., {Barban}, C., {Appourchaux}, T., \&
  {Auvergne}, M. 2009, \aap, 495, 979

\bibitem[{{Michel} {et~al.}(2008){Michel}, {Baglin}, {Auvergne}, {Catala},
  {Samadi}, {Baudin}, {Appourchaux}, {Barban}, {Weiss}, {Berthomieu},
  {Boumier}, {Dupret}, {Garcia}, {Fridlund}, {Garrido}, {Goupil}, {Kjeldsen},
  {Lebreton}, {Mosser}, {Grotsch-Noels}, {Janot-Pacheco}, {Provost},
  {Roxburgh}, {Thoul}, {Toutain}, {Tiph{\`e}ne}, {Turck-Chieze}, {Vauclair},
  {Vauclair}, {Aerts}, {Alecian}, {Ballot}, {Charpinet}, {Hubert},
  {Ligni{\`e}res}, {Mathias}, {Monteiro}, {Neiner}, {Poretti}, {Renan de
  Medeiros}, {Ribas}, {Rieutord}, {Cort{\'e}s}, \& {Zwintz}}]{Michel2008}
{Michel}, E., {et~al.} 2008, Science, 322, 558

\bibitem[{{Molenda-Zakowicz} {et~al.}(2013){Molenda-Zakowicz}, {Sousa},
  {Frasca}, {Uytterhoeven}, {Briquet}, {Van Winckel}, {Drobek}, {Niemczura},
  {Lampens}, {Lykke}, {Bloemen}, {Gameiro}, {Jean}, {Volpi}, {Gorlova},
  {Mortier}, {Tsantaki}, \& {Raskin}}]{Molenda2013arxiv}
{Molenda-Zakowicz}, J., {et~al.} 2013, ArXiv e-prints

\bibitem[{{Montalb{\'a}n} {et~al.}(2010){Montalb{\'a}n}, {Miglio}, {Noels},
  {Scuflaire}, \& {Ventura}}]{Montalban2010}
{Montalb{\'a}n}, J., {Miglio}, A., {Noels}, A., {Scuflaire}, R., \& {Ventura},
  P. 2010, Astronomische Nachrichten, 331, 1010

\bibitem[{{Mosser} {et~al.}(2011{\natexlab{a}}){Mosser}, {Barban},
  {Montalb{\'a}n}, {Beck}, {Miglio}, {Belkacem}, {Goupil}, {Hekker}, {De
  Ridder}, {Dupret}, {Elsworth}, {Noels}, {Baudin}, {Michel}, {Samadi},
  {Auvergne}, {Baglin}, \& {Catala}}]{Mosser2011a}
{Mosser}, B., {et~al.} 2011{\natexlab{a}}, \aap, 532, A86

\bibitem[{{Mosser} {et~al.}(2011{\natexlab{b}}){Mosser}, {Belkacem}, {Goupil},
  {Michel}, {Elsworth}, {Barban}, {Kallinger}, {Hekker}, {De Ridder}, {Samadi},
  {Baudin}, {Pinheiro}, {Auvergne}, {Baglin}, \& {Catala}}]{Mosser2011b}
---. 2011{\natexlab{b}}, \aap, 525, L9

\bibitem[{{Mosser} {et~al.}(2012{\natexlab{a}}){Mosser}, {Elsworth}, {Hekker},
  {Huber}, {Kallinger}, {Mathur}, {Belkacem}, {Goupil}, {Samadi}, {Barban},
  {Bedding}, {Chaplin}, {Garc{\'{\i}}a}, {Stello}, {De Ridder}, {Middour},
  {Morris}, \& {Quintana}}]{Mosser2012b}
---. 2012{\natexlab{a}}, \aap, 537, A30

\bibitem[{{Mosser} {et~al.}(2012{\natexlab{b}}){Mosser}, {Goupil}, {Belkacem},
  {Michel}, {Stello}, {Marques}, {Elsworth}, {Barban}, {Beck}, {Bedding}, {De
  Ridder}, {Garc{\'{\i}}a}, {Hekker}, {Kallinger}, {Samadi}, {Stumpe},
  {Barclay}, \& {Burke}}]{Mosser2012a}
---. 2012{\natexlab{b}}, \aap, 540, A143

\bibitem[{{Osaki}(1975)}]{Osaki1975}
{Osaki}, J. 1975, \pasj, 27, 237

\bibitem[{{Roxburgh}(2009)}]{Roxburgh2009}
{Roxburgh}, I.~W. 2009, \aap, 493, 185

\bibitem[{{Samadi} {et~al.}(2008){Samadi}, {Belkacem}, {Goupil}, {Dupret}, \&
  {Kupka}}]{Samadi2008}
{Samadi}, R., {Belkacem}, K., {Goupil}, M.~J., {Dupret}, M.-A., \& {Kupka}, F.
  2008, \aap, 489, 291

\bibitem[{{Scuflaire}(1974)}]{Scuflaire1974}
{Scuflaire}, R. 1974, \aap, 36, 107

\bibitem[{{Shibahashi}(1979)}]{Shibahashi1979}
{Shibahashi}, H. 1979, \pasj, 31, 87

\bibitem[{{Toutain} \& {Gouttebroze}(1993)}]{Toutain1993}
{Toutain}, T., \& {Gouttebroze}, P. 1993, \aap, 268, 309

\end{thebibliography}

\clearpage

\clearpage

\clearpage

\end{document}